\documentclass[prl,twocolumn,showpacs]{revtex4}

\usepackage{graphicx}
\usepackage{bm}

\begin{document}

\title{Stratified horizontal flow in vertically vibrated granular layers}

\author{M. Levanon}
\author{D. C. Rapaport}
\email{rapaport@bluegum.ph.biu.ac.il}
\affiliation{Physics Department, Bar-Ilan University, Ramat-Gan 52900, Israel}

\date{September 11, 2000}

\begin{abstract}

A layer of granular material on a vertically vibrating sawtooth-shaped base
exhibits horizontal flow whose speed and direction depend on the parameters
specifying the system in a complex manner. Discrete-particle simulations reveal
that the induced flow rate varies with height within the granular layer and
oppositely directed flows can occur at different levels. The behavior of the
overall flow is readily understood once this novel feature is taken into
account.

\end{abstract}


\maketitle

Understanding the static and dynamic properties of granular materials is a
challenging task \cite{jae96} and many aspects of their behavior are
unexpected. An example of this kind of behavior is provided by a granular layer
that is vibrated vertically by a base whose surface profile has a sawtooth
form; in this particular case the material is confined to the annular region
between two upright cylinders and is thus practically two-dimensional. What is
observed \cite{der98,far99}, both in experiment and computer simulation, is
horizontal flow. While this novel effect is perhaps not entirely unexpected
because the shape of the base breaks the horizontal symmetry, the complex
manner in which the flow direction and magnitude depend on the many parameters
that specify the system is surprising. The purpose of this Letter is to
describe a series of simulations that resolve this apparent complexity in terms
of the flows in horizontal strata within the vibrated layer; when this hitherto
unreported aspect of the behavior is taken into account the apparently complex
nature of the overall flow ceases to be a mystery.

The problem of a vibrating granular layer on a flat base is an example of a
granular system where it has proved possible both to validate the simulational
approach based on suitable discrete-particle models and to use simulation to
probe details of the dynamics that are not readily accessible to experiment;
the standing wave patterns that occur in this system have been studied in two
\cite{cle96,lud96,rap98} and three \cite{biz98a,biz98b} dimensions. When the
sinusoidally oscillating base is given a sawtooth surface profile
\cite{der98,far99} horizontal flow appears; experiment and simulation both
display the complex dependence of the flow direction and magnitude on the
parameters defining the system. (While it is also possible to produce
horizontal flow by combining horizontal and vertical base vibration
\cite{gal92}, this is distinct from the sawtooth problem where the vibration is
entirely vertical.) The sawtooth base can be regarded as a kind of ratchet;
evidence of the complexity of systems involving potentials that produce
ratchet-like effects appears in much simpler one-dimensional systems
\cite{doe94,der95}. 


\begin{figure}[b]
\includegraphics[scale=0.34]{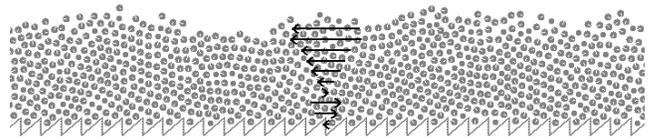}
\caption{Screen snapshot showing a portion of a typical system.}
\label{fig1}
\end{figure}

In the present series of two-dimensional simulations, the interaction
specifying the grain shape is assumed \cite{hir97} to have a Lennard-Jones (LJ)
form, with a cutoff at the range where the repulsive force falls to zero. For
grains located at $\bm{r}_i$ and $\bm{r}_j$ this is $\bm{f}^r_{ij} = (48
\epsilon / r_{ij}) [ (\sigma_{ij} / r_{ij})^{12} - 0.5 (\sigma_{ij} / r_{ij})^6
] \hat{\bm{r}}_{ij}$ for $r_{ij} < 2^{1/6} \sigma_{ij}$ and zero otherwise;
here $\bm{r}_{ij} = \bm{r}_i - \bm{r}_j$ and $\sigma_{ij} = (\sigma_i +
\sigma_j) / 2$. $\sigma_i$ is the approximate diameter of grain $i$, although
since the grains possess a certain degree of softness this is not precisely
defined. The LJ interaction is strongly repulsive at small $r_{ij}$, more so
than linear overlap and Hertzian type repulsions also used in granular
simulation \cite{sch96}. For convenience, we use reduced units in which length
is expressed in terms of the diameter of the largest grains, energy in terms of
$\epsilon$, and a grain with unit diameter (in reduced units) has unit mass.
Grain sizes are randomly distributed between 0.9 and 1.

Normal and transverse viscous damping forces \cite{cun79,wal83} produce the
inelastic collisions and retard sliding during the collision. The normal force
is $\bm{f}^n_{ij} = - \gamma_n (\dot{\bm{r}}_{ij} \cdot \hat{\bm{r}}_{ij})
\hat{\bm{r}}_{ij}$. The transverse force is $\bm{f}^s_{ij} = - {\rm sign}
(v^s_{ij}) \min (\mu | \bm{f}^r_{ij} + \bm{f}^n_{ij} |, \gamma_s | v^s_{ij} |)
\hat{\bm{s}}_{ij}$, where $v^s_{ij} = \dot{\bm{r}}_{ij} \cdot \hat{\bm{s}}_{ij}
+ r_{ij} (\sigma_i \omega_i + \sigma_j \omega_j) / (\sigma_i + \sigma_j)$ is
the relative tangential velocity of the disks, $\hat{\bm{s}}_{ij} = \hat{z}
\times \hat{\bm{r}}_{ij}$ ($\hat{z}$ is the unit normal to the simulation
plane), and $\omega_i$ is an angular velocity. The value of the static friction
coefficient used here is $\mu = 0.5$ and the normal and transverse damping
coefficients are $\gamma_n = \gamma_s = 5$.


\begin{figure}
\includegraphics[scale=0.65]{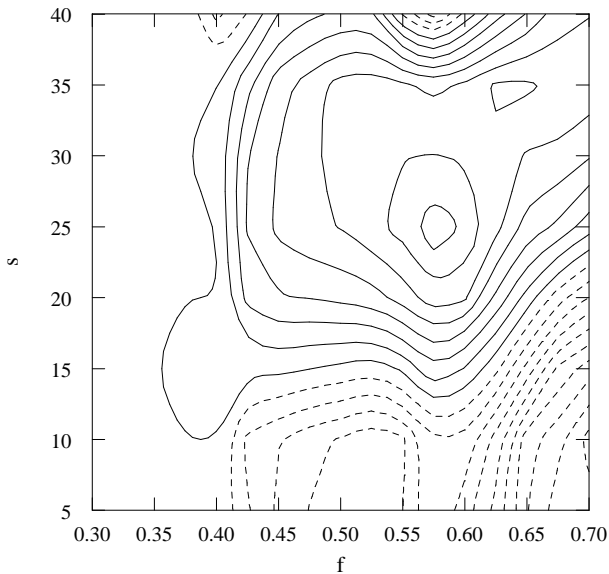}
\caption{Contour plot showing overall horizontal flow velocity as a function of
$f$ and $s$ for $h = 4$ (in reduced units); solid and dashed curves denote
positive and negative values respectively, and the contour interval is 0.013
(the same value is used subsequently).}
\label{fig2}
\end{figure}


\begin{figure}
\includegraphics[scale=0.65]{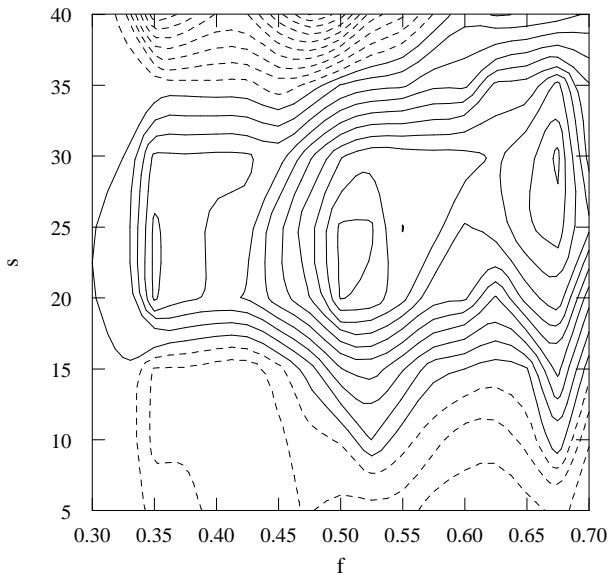}
\caption{Flow velocity (see Fig.~\ref{fig2}) for $h = 10$.}
\label{fig3}
\end{figure}

While most granular simulations are based on particles with a certain amount of
softness, allowing the use of differentiable potentials, there is an
alternative approach involving step potentials; provided the former are not too
soft and the flow involves some degree of fluidization the results of the two
approaches are similar \cite{lud94a,lud94b}. The previous simulations of the
sawtooth problem \cite{far99} used step potentials and the results were limited
to thin layers; the softer potentials of the present work do not restrict the
layer thickness.

The sawtooth base is constructed from a set of grainlike disks positioned to
produce the required profile; these disks oscillate vertically in unison to
produce the effect of a sinusoidally vibrated base. The disks themselves
interact with the grains using the same force laws as above; their diameter is
0.33 and the distance between disk centers is half this value, producing
sawtooth edges that are reasonably straight. The system is horizontally
periodic and extends sufficiently far vertically to prevent grains reaching the
upper boundary. 

Gravity also acts on the grains; here $g = 5$. The total force and torque on
each grain are readily computed and the equations for translational and
rotational motion are numerically integrated using the leapfrog method with a
time step of 0.005. These and other methodological issues are described
elsewhere \cite{rap95}. Each run extends over 1000 base cycles and the results
are grouped into 10 non-overlapping blocks to provide error estimates.

The results described here are for a sawtooth height of 2 and a vibration
amplitude of 1. The sawtooth shape is strongly asymmetric, with a gently
sloping left edge and a very steep right edge; projected onto the base the edge
lengths are in the ratio 99:1. The adjustable parameters are the base vibration
frequency $f$ and the number of sawteeth $s$ ($s$ is specified rather than the
tooth width since it must be an integer). The initial state consists of grains
positioned on a rectangular grid with unit spacing; the horizontal grid size is
fixed at 90 (this is also the system width), and the adjustable vertical grid
size is taken to be the nominal thickness $h$ of the granular layer. 

Fig.~\ref{fig1} shows a screen image (with limited visual resolution) of a
portion of a typical system for the case $s = 40$. Since the base is near its
lowest point the surface waveform is absent. The arrows indicate the stratified
flow discussed below. 


Fig.~\ref{fig2} shows a contour plot of the average horizontal flow velocity as
a function of $f$ (frequency) and $s$ (number of teeth) for layer thickness $h
= 4$. The flow is negative at low $s$, and for certain $f$ also at high $s$
(corresponding to narrow gaps between teeth with space for just a single
grain); horizontal cuts through the plot for certain $s$ values show velocity
reversals as a function of $f$, likewise vertical cuts as a function of $s$.
Fig.~\ref{fig3} is similar but for $h = 10$; the flow is stronger at low $f$
and there is a larger negative region at high $s$. Examination of the entire
range of $h$ values considered (between 1 and 16) reveals smoothly shifting
domain boundaries separating positive and negative flows; merely examining a
few cuts through these two-dimensional plots will miss this gradual variation.
These results include the complex parameter dependence observed previously
\cite{der98,far99}, but with reversed signs because of the opposite sawtooth
asymmetry.

\begin{figure}
\includegraphics[scale=0.65]{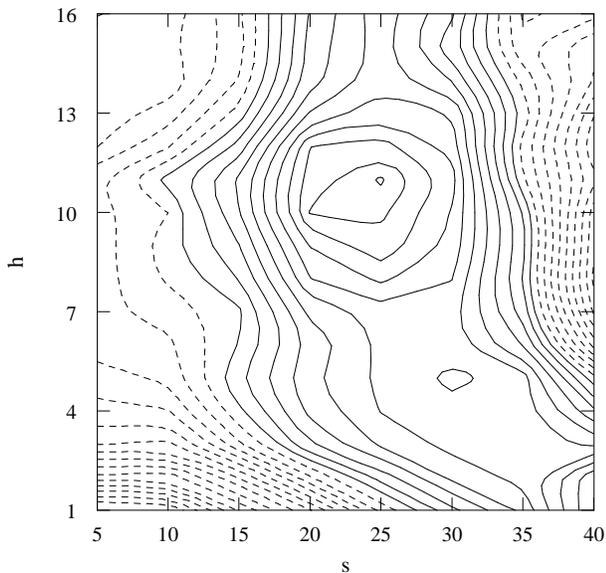}
\caption{Flow velocity as a function of $s$ and $h$ for $f = 0.5$.}
\label{fig4}
\end{figure}

The results shown in Fig.~\ref{fig4} deal with another two-dimensional
cross-section through the multidimensional parameter space and reveal the flow
dependence on $s$ and $h$ for $f = 0.5$. Flow reversals are seen to occur both
as functions of $s$ and $h$ (including monolayers for which $h = 1$). Other
values of $f$ over the range examined (0.3 to 0.7) again show a gradual
shifting of the boundaries between regions of oppositely directed flow.


Space precludes discussion of the corresponding results for other parameter
choices (such as sawtooth height and asymmetry) which are similar in essence
but differ in detail. A particularly drastic change is to set the tangential
damping to zero, thereby removing the rotational component of the dynamics.
Once again the quantitative details change, but the overall behavior is not
strongly affected; the main effect of rotation and the associated damping is to
increase energy dissipation.

Further progress in understanding the behavior requires looking beyond the
overall flow and taking advantage of the simulations to examine the dynamics in
greater detail. An analogy can be made with thermal convection where the mean
flow is zero, but this does not preclude the presence of highly structured
convection rolls. Here, too, there are more localized flow structures that can
only be detected by probing the spatial variation of the flow; Fig.~\ref{fig1}
already hinted at the kind of behavior that can be observed if flows in
distinct horizontal strata are measured. The experiments carried out on this
system \cite{der98,far99} examined just the overall flow by measuring tracer
particle motion, and the accompanying simulations were confined to thin layers
only, so there was little opportunity for observing any spatial variation.

Studying stratified flow requires assigning grains to horizontal levels while
allowing for the fact that grains can migrate vertically. Since this procedure
is not uniquely defined, the approach adopted here is to reassign the grains
once every vibration cycle when the base reaches its lowest point. This has
certain shortcomings, particularly for higher $f$, because grains in the peaks
of the surface waves are assigned to higher levels; however, apart from some
smearing of the measurements between levels, no spurious effects are expected
from this approach.


\begin{figure}
\includegraphics[scale=0.65]{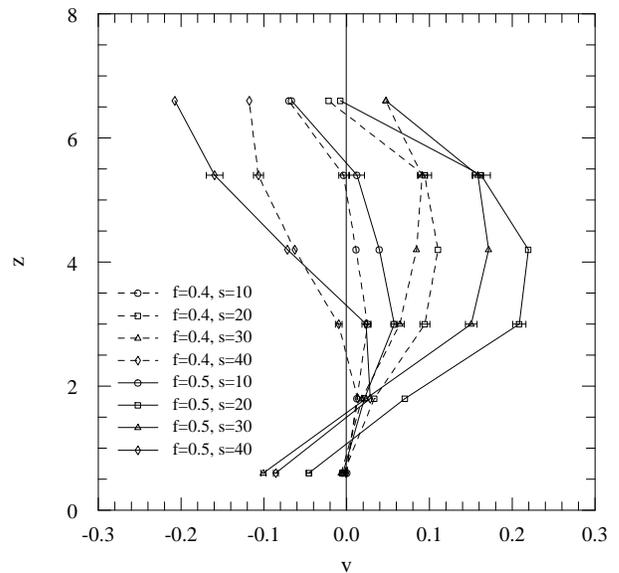}
\caption{Stratified flow velocity ($v$) as a function of height ($z$) above the
base for $h = 8$; typical error bars are shown.}
\label{fig5}
\end{figure}


\begin{figure}
\includegraphics[scale=0.65]{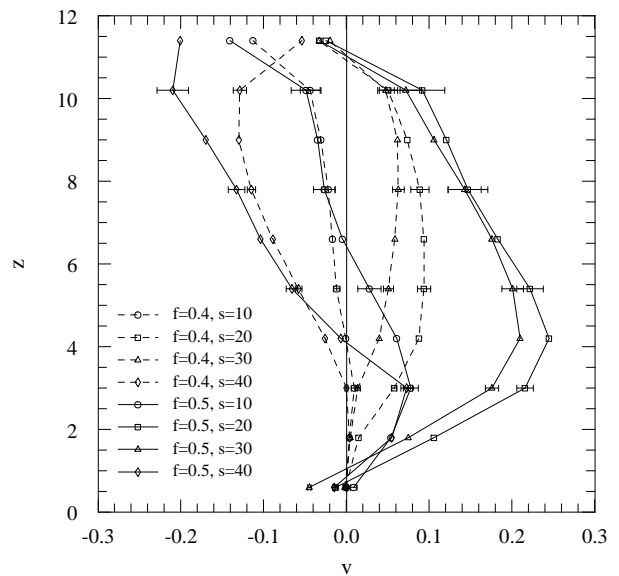}
\caption{Stratified flow velocity (see Fig.~\ref{fig5}) for $h = 12$.}
\label{fig6}
\end{figure}

Fig.~\ref{fig5} shows the height ($z$) dependence of the stratified flow
velocity for $h = 8$, for several sets of $f$ and $s$ values; the thickness of
each level is 1.2, which is slightly larger than the grain size.
Fig.~\ref{fig6} shows the corresponding results for $h = 12$. A typical
velocity profile starts with negative or near-zero flow at the bottom level;
the flow initially increases with $z$, reaches a positive maximum, and then
starts to drop, in most cases becoming negative again. The curves for $s = 10$
(wide teeth) and $s = 40$ (narrow teeth) are the most negative at the upper
levels, while intermediate curves show strong positive maxima near $z = 4$. 


The preferred flow directions at different levels reflect the key features of
the system. The behavior of the grains near the base is reminiscent of a thin
layer that can flow in either direction, depending on the prevailing
conditions. On the other hand, the asymmetry of the sawteeth used here, which
in themselves would be more likely to reflect a single falling grain in the
negative direction, indirectly influences the behavior in the upper levels;
this effect is transmitted through the intervening material which could well be
moving in the opposite direction. When there is positive flow at the lower
levels the competition between these opposing effects produces the
counterflowing velocity profiles. This idealized scheme is of course
complicated by other aspects including grain rotation and the fact that the
grains at the bottom can be permanently trapped between the teeth thereby
reducing the effective sawtooth height.

The explanation for the puzzling reversals in overall flow direction is
embodied in curves such as those shown in Figs.~\ref{fig5} and \ref{fig6}. The
overall flow is the density-weighted mean of the stratified flow velocity; its
sign depends on the form of the velocity profile, itself a consequence of the
relative strengths of the positive and negative flows occurring at different
levels. 

The more familiar role of simulation, particularly as it applies to granular
media, is in attempting to reproduce experimentally observed behavior. In this
Letter the process has been inverted, and through simulation it has proved
possible to identify a previously unknown mechanism in the model granular
system. On the assumption that the model correctly describes the relevant
features of real granular materials this amounts to a prediction that calls for
experimental verification. Should this kind of behavior be found
experimentally, stratified flow would be an interesting addition to the
extensive repertoire of exotic properties exhibited by granular matter.

\begin{acknowledgments}

This research was supported in part by the Israel Science Foundation. We are
grateful to T. Vicsek for introducing us to the problem.

\end{acknowledgments}

\bibliography{gratchet}

\end{document}